# MOEMS deformable mirror testing in cryo for future optical instrumentation

Frederic Zamkotsian[1], Patrick Lanzoni[1], Rudy Barette[1], Emmanuel Grassi[1], Patrick Vors[1]
Michael Helmbrecht[2], Franck Marchis[2], Alex Teichman[2]

[1] Laboratoire d'Astrophysique de Marseille (LAM), CNRS, 38 rue Frederic Joliot Curie, 13388 Marseille Cedex 13, France
[2] Iris AO, 2930 Shattuck Avenue, Berkeley, CA 94705, USA

e-mail: frederic.zamkotsian@lam.fr

**ABSTRACT**

MOEMS Deformable Mirrors (DM) are key components for next generation optical instruments implementing innovative adaptive optics systems, in existing telescopes as well as in the future ELTs. Due to the wide variety of applications, these DMs must perform at room temperature as well as in cryogenic and vacuum environment.
Ideally, the MOEMS-DMs must be designed to operate in such environment. This is unfortunately usually not the case. We will present some major rules for designing / operating DMs in cryo and vacuum.
Next step is to characterize with high accuracy the different DM candidates. We chose to use interferometry for the full characterization of these devices, including surface quality measurement in static and dynamical modes, at ambient and in vacuum/cryo. Thanks to our previous set-up developments, we are placing a compact cryo-vacuum chamber designed for reaching 10-6 mbar and 160K, in front of our custom Michelson interferometer, able to measure performances of the DM at actuator/segment level as well as whole mirror level, with a lateral resolution of 2µm and a sub-nanometric z-resolution.
Using this interferometric bench, we tested the PTT 111 DM from Iris AO: this unique and robust design uses an array of single crystalline silicon hexagonal mirrors with a pitch of 606µm, able to move in tip, tilt and piston with strokes from 5 to 7µm, and tilt angle in the range of +/- 5mrad. They exhibit typically an open-loop flat surface figure as good as < 20nm rms. A specific mount including electronic and opto-mechanical interfaces has been designed for fitting in the test chamber. Segment deformation, mirror shaping, open-loop operation are tested at room and cryo temperature and results are compared. The device could be operated successfully at 160K. An additional, mainly focus-like, 500 nm deformation is measured at 160K; we were able to recover the best flat in cryo by correcting the focus and local tip-tilts on some segments. Tests on DM with different mirror thicknesses (25µm and 50µm) and different coatings (silver and gold) are currently under way.
Finally, the goal of these studies is to test DMs in cryo and vacuum conditions as well as to improve their architecture for staying efficient in harsh environment.

## 1. INTRODUCTION

Several research groups around the world are currently involved in the design of highly performing adaptive optical (AO) systems as well as for next generation instrumentation of 10m-class telescopes than for future extremely large optical telescopes.

Wavefront correction like adaptive optics systems are based on a combination of three elements, the wavefront sensor for the measurement of the shape of the wavefront arriving in the telescope, the deformable mirror is the correcting element, and finally the real time computer closing the loop of the system at a frequency ranging from 0.5 to 3kHz, in order to follow the evolution of the atmospherical perturbations (Fig. 1).

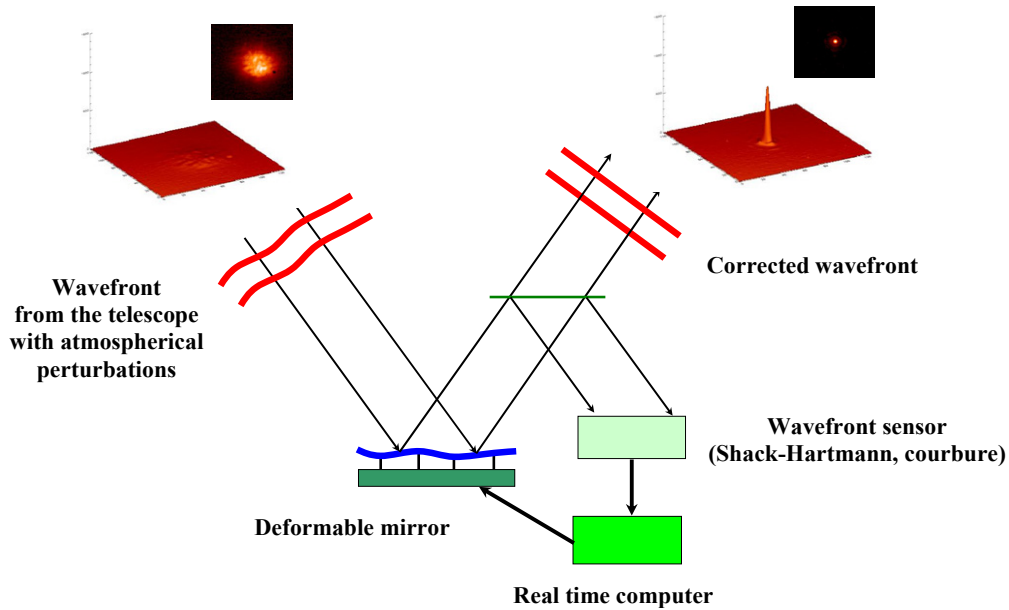

Fig. 1: Schematic of a wavefront correction system

Four main types of AO systems have been built or are under development: Single-Conjugate Adaptive Optics (SCAO), Multi-Conjugate Adaptive Optics (MCAO), Multi-Object Adaptive Optics (MOAO), and Extreme Adaptive Optics (ExAO). These AO systems are associated with different types of WaveFront Sensors (WFS), combined with natural guide stars or laser guide stars, and different architectures of Deformable Mirrors (DM). Numerous science cases will use these AO systems, SCAO, the "classical" AO system will provide accurate narrow field imagery and spectroscopy, MCAO, wide field imagery and spectroscopy, MOAO, distributed partial correction AO, and high dynamic range AO for the detection and the study of circumstellar disks and extra-solar planets. Corrected fields will vary from few arcsec to several arcmin.

These systems require a large variety of deformable mirrors with very challenging parameters. For a 8m telescope, the number of actuators varies from a few 10 up to 5000; these numbers increase impressively for a 40m telescope, ranging from a few 100 to over 50 000, the inter-actuator spacing from less than 200 μm to 1 mm, and the deformable mirror size from 10 mm to a few 100 mm. Conventional technology cannot provide this wide range of deformable mirrors. The development of new technologies based on micro-opto-electro-mechanical systems (MOEMS) is promising for future deformable mirrors. The major advantages of the micro-deformable mirrors (MDM) are their compactness, scalability, and specific task customization using elementary building blocks. This technology permits the development of a complete generation of new mirrors. However this technology has also some limitation. For example, pupil diameter is an overall parameter and for a 40 m primary telescope, the internal pupil diameter cannot be reduced below 0.5 m. According to the maximal size of the wafers (8 inches), a deformable mirror based on MOEMS technology cannot be build into one piece. New AO architectures have been proposed to avoid this limitation. [1]

LAM is involved since several years in conception of new MOEMS devices as well as in characterization of these components for the future instrumentation of ground-based and space telescopes. These studies include programmable slits for application in multi-object spectroscopy (JWST, European networks, EUCLID, BATMAN), deformable mirrors for adaptive optics, and programmable gratings for spectral tailoring.

We are particularly engaged in a European development of micromirror arrays (MMA) called MIRA for generating reflective slit masks in future Multi-object spectroscopy (MOS) instruments; this technique is a powerful tool for space and ground-based telescopes for the study of the formation and evolution of galaxies. MMA with 100 x 200 μm$^2$ single-crystal silicon micromirrors were successfully designed, fabricated and tested. Arrays are composed of 2048 micromirrors (32 x 64) with a peak-to-valley deformation less than 10 nm, a tilt angle of 24° for an actuation voltage of 130 V. The micromirrors were actuated successfully before, during and after cryogenic cooling, down to

162K. The micromirror surface deformation was measured at cryo and is below 30 nm peak-to-valley. [2,3] In order to fill large focal planes (mosaicing of several chips), we are currently developing large micromirror arrays integrated with their electronics.

LAM is also leading the conception and realization of new MOEMS-based instruments. We are developing a 2048x1080 Digital-Micromirror-Device-based (DMD) MOS instrument to be mounted on the Telescopio Nazionale Galileo (TNG) and called BATMAN. A two-arm instrument has been designed for providing in parallel imaging and spectroscopic capabilities. BATMAN on sky is of prime importance for characterizing the actual performance of this new family of MOS instruments, as well as investigating new observational modes on astronomical objects, from faint and remote galaxies to active areas in nearby galaxies and small bodies of the solar system. This instrument will be placed at TNG by end-2017. [4]

In this paper, we present the specific set-up for the interferometric characterization of a segmented deformable mirror from Iris AO. The results on the mirror surface continuously measured from ambient down to 160K are shown and analyzed.

## 2. DEFORMABLE MIRRORS

Three main Micro-Deformable Mirrors (MDM) architectures are under study in different laboratories and companies. First, the bulk micro-machined continuous-membrane deformable mirror, studied by Delft University and OKO company, is a combination of bulk silicon micromachining with standard electronics technology[5]. This mirror is formed by a thin flexible conducting membrane, coated with a reflective material, and stretched over an electrostatic electrode structure. This mirror shows a very good mirror quality, but the mean deformed surface is a concave surface, and the number of actuators cannot be scalable to hundreds of electrodes. Second, the segmented, micro-electro-mechanical deformable mirror realized by Iris AO [6] consists of a set of segmented piston-tip-tilt moving surfaces, fabricated in dense array. For adaptive optics application, the wavefront has to be properly sampled, increasing the number of actuators for a given number of modes to be corrected. Third, the surface micro-machined continuous-membrane deformable mirror made by Boston Micromachines Corporation (BMC) is based on a single compliant optical membrane supported by multiple attachments to an underlying array of surface-normal electrostatic actuators [7]. The efficiency of this device has been demonstrated recently in several AO system, including the GPI instrument on Gemini telescope. The third concept is certainly the most promising architecture, but it shows limited strokes for large driving voltages, and mirror surface quality may need further improvement for Extreme AO. All these devices are based on silicon or polysilicon materials.

### 2.1. BMC Deformable Mirror

BMC produces the most advanced MEMS deformable mirrors. The concept is based on an array of electrostatic actuators linked one by one to a continuous top mirror (Fig. 2). Their main parameters are approaching the requirements values, i.e. large number of actuators (up to 4096, see Fig. 2), large stroke (up to 5.5 µm), good surface quality, but they still need large voltages for their actuation (150–250V).

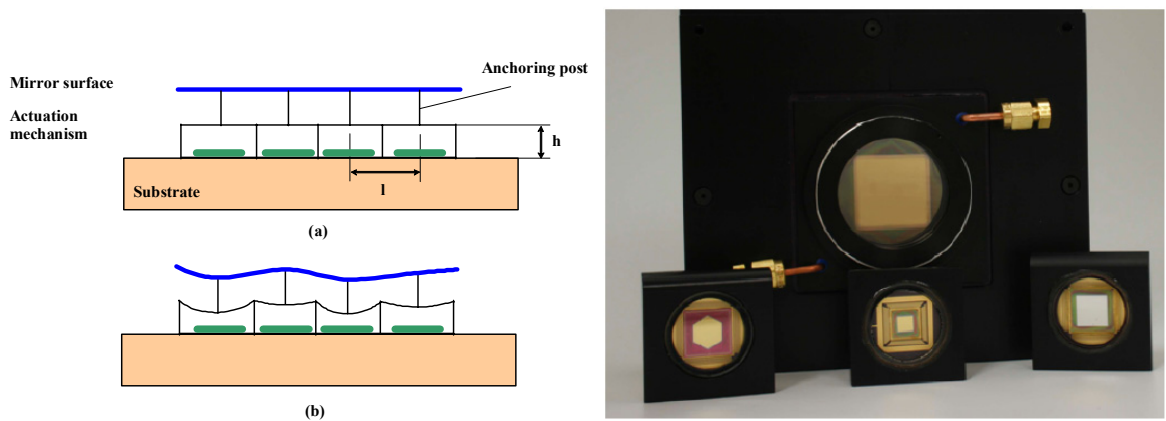

Fig. 2: Continuous membrane MDM from BMC

## 2.2. Iris AO Deformable Mirror

IRIS AO is producing segmented piston tip tilt mirrors with very flat mirrors. An exploded-view schematic diagram is presented in Fig. 3a. The DM array is paved by 37 hexagonal segments with a size of 700 μm from vertex-to-vertex, with a 606μm pitch. The segment is capable of moving in piston/tip/tilt motions (PTT). In Fig. 3a, scaling is highly exaggerated in the vertical direction. In Fig. 3b is shown a die photograph of a 111-actuator 37-piston/tip/tilt-segment DM with 3.5 mm inscribed aperture (PTT111 device). The DM is manufactured using typical MEMS and integrated circuit materials such as polycrystalline silicon (polysilicon), silicon dioxides, silicon nitrides, and a proprietary bimorph material with similar coefficient of thermal expansion (CTE) to that of polysilicon. The s-shape of the bimorph flexures that elevates the DM segment is a result of engineered residual tensile stresses in the bimorph and actuator-platform polysilicon. After the DM is fabricated using highly stable MEMS materials, it is mounted onto a ceramic pin-grid array (PGA) package using an epoxy. The DM is sealed in nitrogen by epoxying a cover window over the DM.

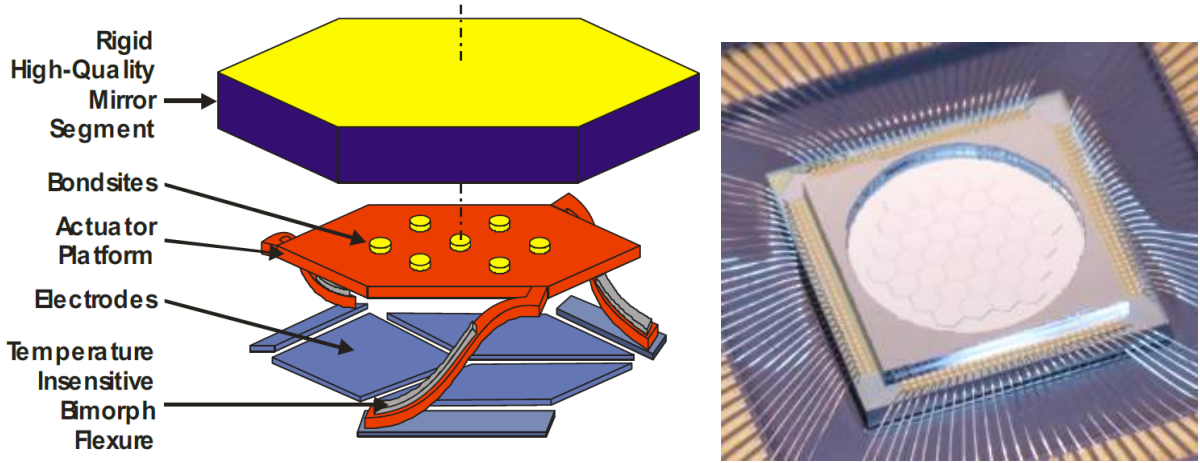

Fig. 3: Segmented MDM from Iris AO (a) Concept of one segment of the mirror;
(b) Die photograph of a 111-actuator 37-piston/tip/tilt-segment DM with 3.5 mm inscribed aperture (PTT111 device)

To actuate the DM, the red actuator-platform layer is held at ground potential and the three diamond-shaped electrodes are energized at different electrical potentials. Applying the same voltage to all three electrodes pulls the segment in a piston motion toward the electrodes. A differential voltage across the electrodes results in tip and tilt motions. Because the positioning is highly repeatable, the DM segment motion can be calibrated, thus linearizing the DM position into orthogonal coordinates.

## 3. CRYOGENIC INTERFEROMETRIC TEST SET-UP

The Laboratoire d'Astrophysique de Marseille has developed over the last few years an expertise in the characterization of micro-optical components. Our expertise in small-scale deformation characterization on the surface of micro-optical components has been conducted initially within the framework of the NASA study of a multi-object infrared spectrograph equipped with MOEMS-based slit masks for the JWST.

### 3.1. Interferometric bench

An interferometric characterization bench has been developed in order to measure the shape and the deformation parameters of these devices. All optical characterizations in static or dynamic behavior are performed, including measurements of optical surface quality at different scales, actuators stroke, maximum mirror deformation and cut-off frequency. This bench is a high-resolution and low-coherence Twyman-Green interferometer (Fig. 4) The light source is an halogen lamp with an interference filter (typical example: $\lambda_0$=650nm, $\Delta\lambda$=10nm). Monitoring the temporal light coherence, this illumination avoids all extraneous fringes induced by classical high coherence sources such as lasers. Conceived as a modular bench, a simple lens change offers two magnification configurations: (1) high in-plane resolution or (2) large field of view authorizing either a very sharply (around 4μm) analysis of the micro-mirror structure inside a small field (typically 1mm), or the whole device study with larger size (up to 40mm). For the high-resolution configuration, diffraction limit (N=3) is reached. **Out-of-plane measurements are performed with phase-shifting interferometry showing very high resolution (standard deviation<1nm)**. A picture of the bench is shown in Fig. 5, mounted on a damped optical table, and surrounded by a Plexiglas enclosure. Features such as optical quality or electro-mechanical behavior are extracted from these high precision three-dimensional component maps. Range is increased without loosing accuracy by using two-wavelength phase-shifting interferometry authorizing large steps measurements. Dynamic analysis like vibration mode and cut-off frequency is also measured with time-averaged interferometry. [8]

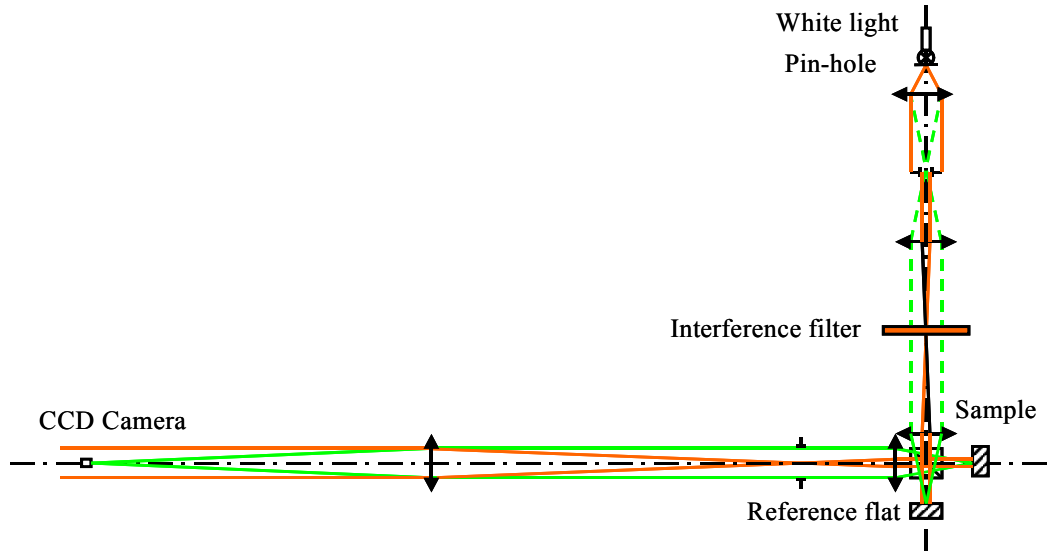

Fig. 4: Schematic description of our interferometric measurement bench.

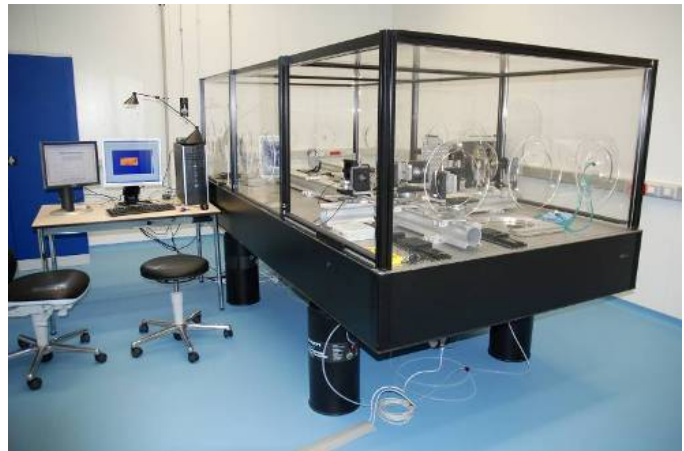

Fig. 5: Picture of the experimental set-up.

### 3.2. Cryogenic experiment

Cryogenic characterization was carried out in a custom built cryogenic chamber installed in front of our interferometric setup. The cryo-chamber has a pressure as low as 10e-6 mbar and is able to cool down to 100K, when not loaded, using a cryogenic generator. In order to get such temperature, the chamber is equipped with an internal screen insulating radiatively the sample from the chamber. Control of the environment is obtained by means of temperature sensors and local heaters. They are wired to the outside environment through a Dutch connector and connected to a custom built control electronics and control-command software.

The chamber has a glass window that allows observing and measuring the sample chip during cryogenic testing. The presence of a glass window at the entrance of the chamber is an issue for getting fringes with a high contrast. Two elements have to be corrected:
- The path difference between the interferometer arms (sample arm and reference mirror arm).
- The glass medium is dispersive for the different wavelengths, each wavelength following a different path.

The first point could be overcome by moving the reference mirror in order to balance the path difference induced by the index difference between the window material and air, this balance is obtained for a very narrow linewidth. As we are using sources with low coherence, i. e. with a wide linewidth (typically 10nm), and as glass is dispersive, the path followed by each wavelength will be slightly different, degrading drastically the fringe pattern contrast. The only solution is to introduce in the reference arm a glass plate exactly identical to the window in the sample arm. In our set-up, the chamber window and the compensation plate have been manufactured at the same time in the same glass material. [9]

### 3.3. PTT111 mounting

The PTT111 device is packaged in PGA chip carrier. The PGA is inserted in a ZIF-holder integrated on a PCB board. Large metallic surfaces on the PCB facilitate cooling down the system; renouncing the solder-stop layer eases outgassing of the PCB base material during evacuation of the chamber. The PCB itself is mounted via a fix-point-plane-plane attachment system to a solid aluminum block, the latter being interconnected to the cryo-generator (Fig. 6a). Thick copper wires between the PCB and the aluminum block further enhance thermal transport between the sample chip and the cryostat. Two ribbons cables allow interconnecting up to 120 electrical wires through two 100-pins and 20-pins feed-throughs. On the outside environment the wires are connected to the Iris AO control electronics. Temperature sensors are connected to the aluminum block and to window frame on top of the device.

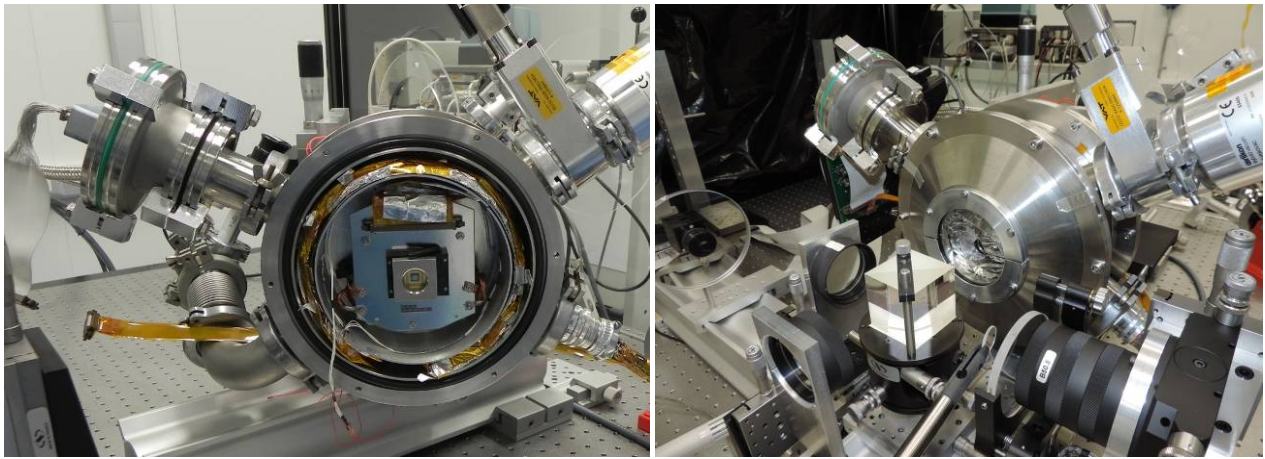

Fig. 6: (a) PTT111 device mounted in the cryogenic chamber for characterization in space environment;
(b) Front window is closed and the cryogenic chamber is installed in front of our interferometric setup
The segmented deformable mirror could be successfully actuated before, during and after cryogenic cooling at 160K.

The chamber is then closed by a flange and placed in front of the interferometer (Fig. 6b). Along the reference path, two compensation plates are placed for compensating the chamber window (large plate) and the device window (small plate). By this way, we keep a high contrast for the interferometric fringes.

## 4. PTT111 SURFACE CHARACTERIZATION

Our experiment is done on an engineering grade device where the segments #23 and #24 are lockouts. The segment thickness is 25μm and the coating is protected silver. The maximum array stroke is 3.01μm, and the maximum tilt angle is 5mrad. Fig. 7 is a picture of the device made on our bench (without the interferometric fringes). The two lockouts segments are at the upper right.

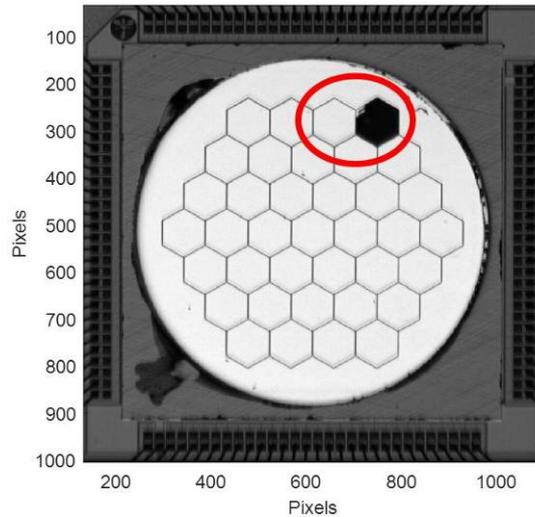
Fig. 7: PTT111 device tested in our experiment (engineering grade device)
Segments #23 and #24 (in the red circle) are lockouts.

The device is driven either by the *Graphical User Interface* (GUI) provided by Iris-AO for the integration and pre-characterization phases. The interferometric measurements are done with the LAM-developed software, in Matlab, and linked with the Matlab driver provided by Iris AO. The GUI is showing a view of the mirror with numbered segments, and global Zernike coefficient as well as local (at segment level) Zernike coefficients could be tuned for each actuator/segment.

A calibration step has been done by Iris AO for measuring each actuator response (3 actuators /segment). Then, a "best flat" condition is calculated in order to minimize the residual wavefront error on the surface, and applied. In Fig. 8 is a screen shot made at the beginning of the experiment when the best flat condition is applied; at left hand side, the interferometric image of the mirror with "best flat" condition; at right bottom, view of the Iris AO GUI for driving the device.

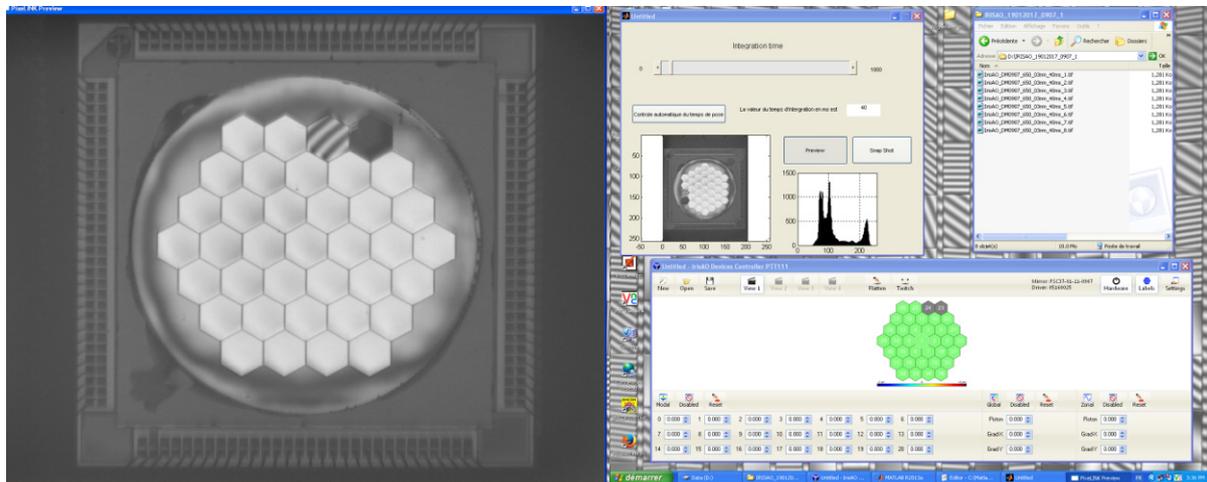
Fig. 8: Screen shot during the experiment; left: interferometric image of the mirror when the "best flat" condition is applied; right bottom: Iris AO GUI for driving the device.

We could then apply a serial of commands on the mirror. In Fig. 9, different mirror configurations are presented. From top left to bottom right, we can see:
- a pure piston (150nm) on the central M1 segment,
- a global astigmatism on the mirror, using the global Zernike coefficient set at 0.125
- a serial of three identical tilts on all segments in X direction, with 0.25mrad, 1.5mrad and 4.9mrad respectively.

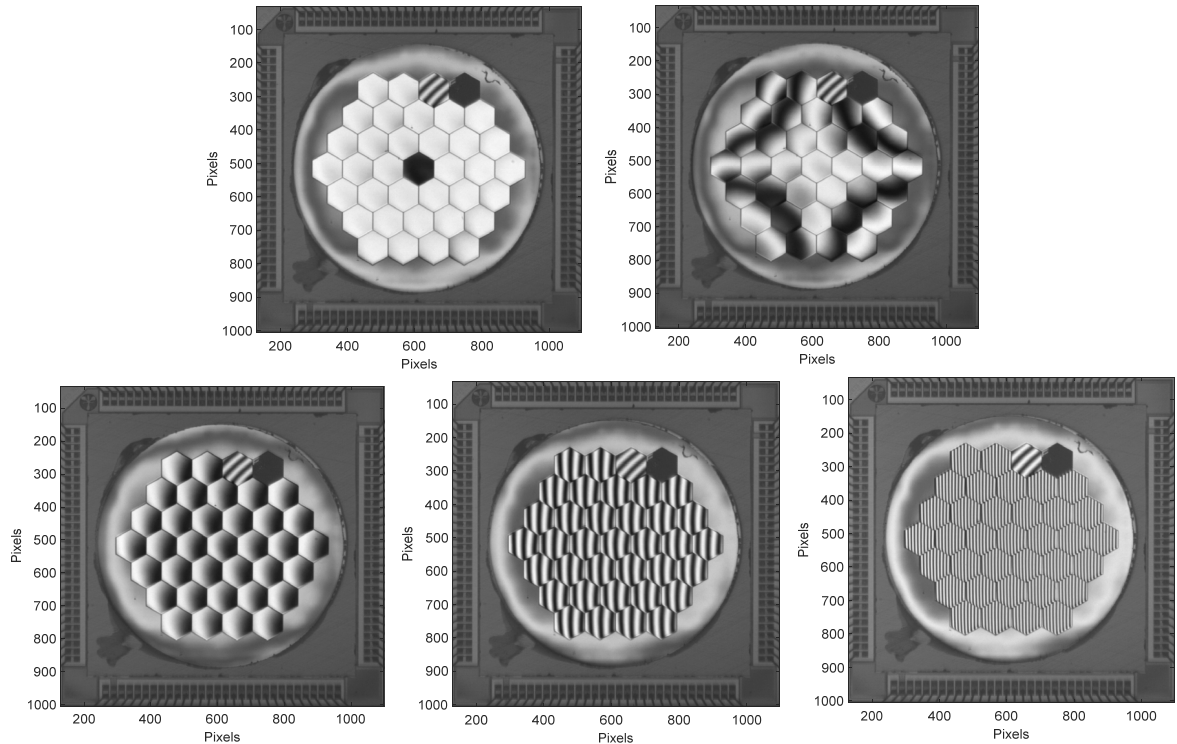

Fig. 9: top left: 150nm piston on segment#1; top right: astigmatism (0.125 on Zernike coefficient);
Bottom: Increasing tilt values along the X direction are applied to all segments (0.25, 1.5 and 4.9 mrad).

### 4.1. Best flat at ambient

From the interferometric measurement, we obtain the surface deformation shown in Fig. 10. **The best flat residual over the whole mirror is very good with 13.5nm RMS, and 110.2nm Peak-to-Valley (PtV).**
This result shows the high quality of the mirror architecture and of the fabrication process. This flatness is a combination of a very good reproducibility of the actuator platform position after his elevation thanks to the bimorph flexures (Fig. 3a), and the choice of thick single-crystalline Silicon for the segment material. This position is very stable (long term measurement not done yet on position stability and reproducibility).

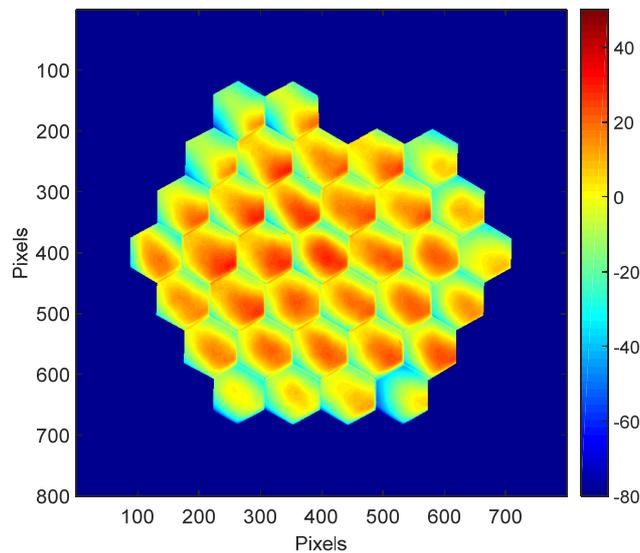

Fig. 10: Best flat surface deformation at ambient (13.5nm RMS, 110.2nm PtV).

At segment level, the residual wavefront error is with a slight convex shape, observed on most of the segments.

### 4.2. Best flat at cryo

The device is then cooled down slowly from ambient temperature (293K) down to 160K, with the device constantly operating in its best flat condition. **The PTT111 device is operating properly at all temperatures between 293K and 160K, and in vacuum.**

Every 10K an interferometric measurement is done in order to follow the differential deformation of the mirror at whole mirror level as well as at segment level. Several patterns are applied and measured in order to see the ability of the device to behave as at room temperature; the applied "patterns" are best flat, pure pistons on some segments, and different tilts on the segments. Due to the vibrations induced by the cryo pump on the sample, we have to stop it during the measurement, leading to a limited increase of the temperature during the measurement duration. Phase shifting interferometry parameters have been adjusted in order to minimize the measurement time to few minutes.

In Fig. 11a, the best flat surface deformation at cryo (160K) is given with the original best flat as calibrated at ambient. **A global convex deformation is observed reaching a deformation of 86.9nm RMS, 502nm PtV. Some additional deformations (mainly tilts) are observed on some segments at the upper left side (segments # 26, 27 and 28).**

This limited differential deformation is due to the residual CTE mismatch between the structural layers: the actuator platform, the bonding pads, the mirror segment, the coating (see Fig. 3a). The behaviour of the underlying bimorph structure may also participate to this effect.

While the device is still at cold temperature, a corrected best flat is done "by hand" for minimizing the wavefront error. First of all, a global focus coefficient is applied to decrease the global convex bow, and tilts (in X and Y) are adjusted on segments # 26, 27 and 28. **The "corrected best flat" is given in Fig. 11b, with an improved surface deformation, dividing its value by a factor 2, down to 40.4nm RMS, 244nm PtV.**

The mirror is operating perfectly in cryo, and a "new" best flat could then be calibrated at each temperature. Procedures for optimizing these best flats will be developed and applied for further tests.

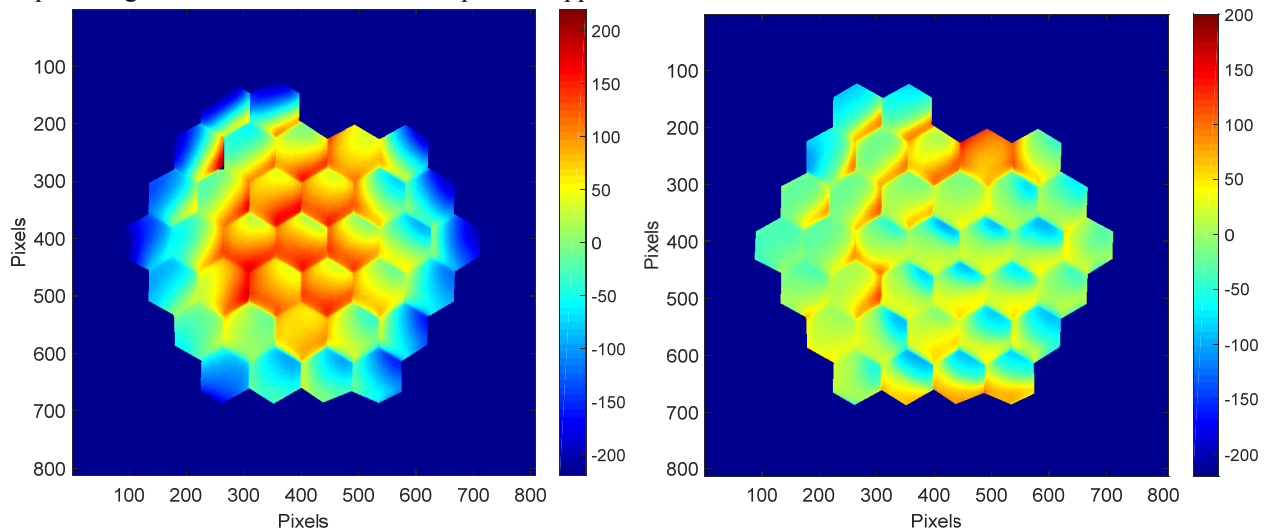

Fig. 11: Best flat surface deformation at cryo (160K)
(a) **Original best flat** as calibrated at ambient (86.9nm RMS, 502nm PtV);
(b) **Corrected best flat** after global focus and segment tilts adjustment (40.4nm RMS, 244nm PtV).

### 4.3. Segment characterization

Thanks to our set-up spatial resolution, we have several thousand measurement points per segment. It is then possible to measure, at segment level, the deformation induced by the strong temperature change from ambient to cryo. We did our preliminary measurement and analysis on the central segment (segment #1).

**In Fig. 12, the DM central segment (segment #1) surface deformation is shown for 3 temperatures, at ambient (293K), at 223K and at 160K, they are respectively of 10.8nm RMS, 47.2nm PtV; 8.4nm RMS, 40.5nm PtV; 8.5nm RMS, 59.8nm PtV.**

We cannot measure any degradation of the segment surface quality when cooling down at cryogenic temperature, as the RMS and PtV values are in the same range, even if at 223K, the surface quality seems to be slightly better. This could be explained by the fact the minimal stress point of the stack of layers/materials of the DM architecture has been reached. However, the deformation value is not the only parameter; if we observe the segment shape, we can clearly see that the convex shape at ambient is changing to an astigmatic shape at cryo. This analysis must be confirmed by observing all segments of the device and scroll over all temperatures.

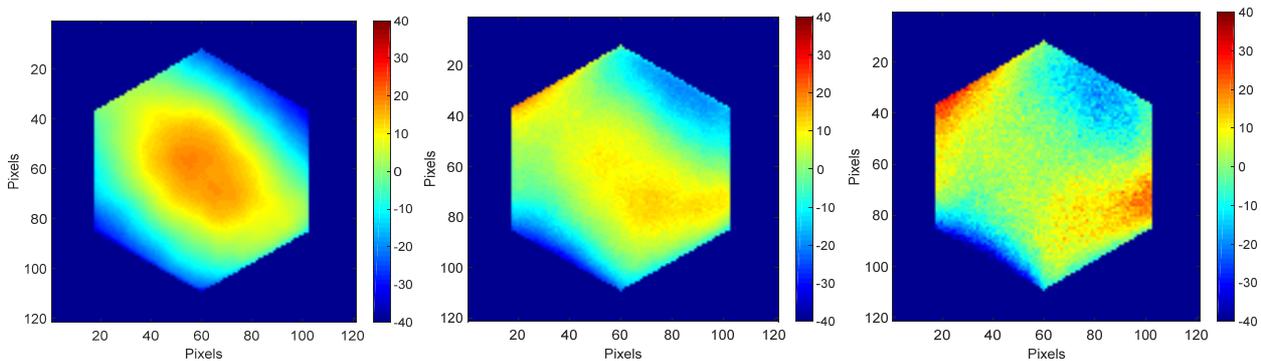

Fig. 12: DM central segment (segment #1) surface deformation
(a) At ambient (293K), 10.8nm RMS, 47.2nm PtV; (b) At 223K 8.4nm RMS, 40.5nm PtV; (c) At 160K, 8.5nm RMS, 59.8nm PtV.

## 5. CONCLUSION

Innovative adaptive optics systems in existing telescopes as well as in the future ELTs need efficient MOEMS Deformable Mirrors (DM) able to perform at room temperature as well as in cryogenic and vacuum environment.

Using a specific interferometric bench coupled with a cryo-vacuum chamber, a PTT 111 DM from Iris AO has been successfully tested from ambient temperature to 160K. The device is properly operating in cryo, revealing an additional, mainly focus-like, 500 nm deformation at 160K; we were able to recover the best flat in cryo by correcting the focus and local tip-tilts on some segments. The segment shape evolution in cryo has also been measured, going from a convex to an astigmatic shape. This analysis must be confirmed by observing all segments of the device and scroll over all temperatures.

Tests on DM with different mirror thicknesses (25µm and 50µm) and different coatings (silver and gold) are currently under way.

Finally, the goal of these studies is to test DMs in cryo and vacuum conditions as well as to improve their architecture for staying efficient in harsh environment.

# REFERENCES

[1] F. Zamkotsian, K. Dohlen, "Prospects for MOEMS-based adaptive optical systems on extremely large telescopes", in *Proceedings of the conference Beyond conventional Adaptive Optics*, Venice, Italy (2001)

[2] S. Waldis, F. Zamkotsian, P.-A. Clerc, W. Noell, M. Zickar, N. De Rooij, "Arrays of high tilt-angle micromirrors for multiobject spectroscopy, " IEEE Journal *of Selected Topics in Quantum Electronics* **13**, pp. 168–176 (2007).

[3] M. Canonica, F. Zamkotsian, P. Lanzoni, W. Noell, N. de Rooij, "The two-dimensional array of 2048 tilting micromirrors for astronomical spectroscopy," Journal of Micromechanics and Microengineering, 23 055009, (2013)

[4] Frederic Zamkotsian, Harald Ramarijaona, Manuele Moschetti, Patrick Lanzoni, Marco Riva, Nicolas Tchoubaklian, Marc Jaquet, Paolo Spano, William Bon, Romain Alata, Luciano Nicastro, Emilio Molinari, Rosario Cosentino, Adriano Ghedina, Manuel Gonzalez, Walter Boschin, Paolo Di Marcantonio, Igor Coretti, Roberto Cirami, Filippo Zerbi, Luca Valenziano, "Building BATMAN: a new generation spectro-imager on TNG telescope ", in Proceedings of the SPIE conference on Astronomical Instrumentation 2016, Proc. SPIE **9908**, Edinburgh, United Kingdom, (2016)

[5] G. Vdovin, S. Middelhoek and P. M. Sarro, "Technology and applications of micromachined silicon adaptive mirrors", Opt. Eng., **36** (5), 1382-1390 (1997)

[6] M.A. Helmbrecht, M. He, C.J. Kempf, F. Marchis "Long-term stability and temperature variability of Iris AO segmented MEMS deformable mirrors", Proc. SPIE **9909**, (2016)

[7] S. Cornelissen, T. G. Bifano "Advances in MEMS deformable mirror development for astronomical adaptive optics", in Proceedings of the SPIE conference on MOEMS 2012, Proc. SPIE **8253**, San Francisco, USA (2012)

[8] A. Liotard, F. Zamkotsian, "Static and dynamic micro-deformable mirror characterization by phase-shifting and time-averaged interferometry", in *Proceedings of the SPIE conference on Astronomical Telescopes and Instrumentation 2004*, Proc. SPIE **5494**, Glasgow, United Kingdom (2004)

[9] F. Zamkotsian, E. Grassi, S. Waldis, R. Barette, P. Lanzoni, C. Fabron, W. Noell, N. de Rooij, " Interferometric characterization of MOEMS devices in cryogenic environment for astronomical instrumentation," Proc. SPIE **6884**, San Jose, USA, (2008)